\documentclass[sigconf]{acmart}

\newtheorem{definition}{Definition}

\usepackage{balance}
\usepackage{listings}
\usepackage{amsmath}
\usepackage{booktabs}
\usepackage{algpseudocode}
\usepackage{multirow}
\usepackage{soul}
\usepackage{url}
\usepackage{xfrac}
\usepackage{graphicx}
\usepackage{subfig}
\usepackage{tabularx}
\usepackage{pifont}
\usepackage{enumitem}
\usepackage{setspace}

\usepackage{adjustbox}
\usepackage{diagbox}
\usepackage{arydshln}

\setcopyright{rightsretained}

\begin{document}
\title{Still Haven't Found What You're Looking For - Detecting the Intent of Web Search Missions from User Interaction Features}



\author{Ran Yu}
\affiliation{%
  \institution{University of Bonn}
  \city{Bonn} 
  \country{Germany} 
  \postcode{53115}
}
\email{ran.yu@uni-bonn.de}

\author{Limock}
\affiliation{%
  \institution{Leibniz University Hannover}
  \city{Hannover} 
  \state{Germany} 
  \postcode{30167}
}
\email{trovato@corporation.com}

\author{Stefan Dietze}
\affiliation{%
  \institution{GESIS - Leibniz Institute for the Social Sciences}
  \city{Cologne} 
  \country{Germany} 
  \postcode{50667}
}
\email{stefan.dietze@gesis.org}


\begin{abstract}
Web search is among the most frequent online activities. Whereas traditional information retrieval techniques focus on the information need behind a user query, previous work has shown that user behaviour and interaction can provide important signals for understanding the underlying intent of a search mission. An established taxonomy distinguishes between transactional, navigational and informational search missions, where in particular the latter involve a learning goal, i.e. the intent to acquire knowledge about a particular topic. We introduce a supervised approach for classifying online search missions into either of these categories by utilising a range of features obtained from the user interactions during an online search mission. Applying our model to a dataset of real-world query logs, we show that search missions can be categorised with an average F1 score of 63\% and accuracy of 69\%, while performance on informational and navigational missions is particularly promising (F1>75\%). This suggests the potential to utilise such supervised classification during online search to better facilitate retrieval and ranking as well as to improve affiliated services, such as targeted online ads. 
\end{abstract}

%
%



\maketitle

\section{Introduction}


Searching the Web is among the most frequent online activities and has become a ubiquitous task. As is common search practice, a coherent search mission, involving a particular search intent, usually involves several queries as well as one or more search sessions (cf. ~\cite{jones2008beyond,hagen2013} and Section \ref{sec:terms}). 


Broder categorized web search missions into having either \textit{navigational}, \textit{transactional} or \textit{informational} intents~\cite{broder2002}. Whereas transactional search missions usually aim at conducting a specific online transaction, such as, purchasing a ticket, navigational queries merely are aimed at leading the user to a dedicated website. In contrast, informational missions imply the intent of a user to acquire some information assumed to be present on one or more web pages. In this context, the same query, for instance, \textit{Elbphilharmonie} may be used to either buy tickets in a particular concert venue (transactional), to reach the Website \url{https://www.elbphilharmonie.de} (navigational) or to acquire knowledge about the \textit{Elbphilharmonie} (informational). 

Whereas traditional information retrieval techniques focus on understanding the information need of a user in order to retrieve and rank search results according to their relevance to the information need, the actual intent behind a search mission has a strong impact on the relevance of documents to the user. However, as documented by the example above, considering the search intent requires to take into account signals beyond the user query, in particular, to consider the user interactions and behavior observable during the online search process. 

Recent work has aimed at addressing this task by applying supervised~\cite{kang2003,lee2005,liu2006,yates2006,jansen2008,kathuria2010} or unsupervised~\cite{yates2006} models on a set of features extracted from the search activity log corresponding to a single query. Features are extracted from multiple dimensions, such as query term, anchor text, SERP click, browsing behavior and Web document. However recent studies have shown that information seeking tasks have grown more sophisticated~\cite{jones2008} and often require one or more queries across multiple search sessions~\cite{kotov2011,liu2010,agichtein2012}.

In contrast to previous work, usually focused on single query search, we recognise that search missions often spread across several queries as well as search sessions. In this work, we aim at automatically classifying the search intent of actual search missions~\cite{hagen2013} into the aforementioned categories, by utilising features observed from user interactions and behavior during the search process, that is, features reflecting the observed queries, mission and browsing behavior. We build supervised models and apply our approach to a corpus of real-world query logs to experimentally evaluate the classification performance. We show that search missions can be categorised with an average F1 score of 63\% and accuracy of 69\% while performance on informational and navigational missions is particularly promising (F1>75\%). Key contributions of this work include (i) a supervised model for the classification of search missions from behavioral features and (ii) the experimental evaluation of a set of features for search intent classification.

Our results suggest potential to utilise such supervised classification during online search in order to better facilitate retrieval and ranking as well as to improve affiliated services, such as targeted online ads. In particular, while recent work has shown that online search behavior correlates with particular learning intents during informational search missions~\cite{gadiraju2018chiir}, detection of the actual user intent is crucial in order to support users with their actual learning needs. 

The paper is organised as follows. In Section 2, we introduce related work, followed by the definition of important concepts used in this work and formal problem statement in Section 3. We present and motivate the user interaction features and used classifiers in Section 4, whereas the experimental setup and evaluation results are presented in Sections 5 and 6. We discuss further findings and insights gained from this work in Section 7 and conclude the paper in Section 8.

\section{Related Work}
In this section we review related literature on automated search intent classification in the context of Web search.

Early studies on intent classification relied on manual approaches, for instance, by asking users through surveys~\cite{broder2002} or by manual annotation of intents through judges~\cite{rose2004}. 

However, while this process does not scale well to large datasets, automatic classification approaches have been explored. Kang et al.~\cite{kang2003} attempted to classify user queries into three categories: the topic-relevant task, the homepage-finding task and the service-finding task. 
They proposed to use linear regression with 4 features extracted from the query terms and anchor text for the query classification. This approach achieved 91.7\% precision and 61.5\% recall. 

Lee et al.~\cite{lee2005} proposed to use linear regression for the classification of informational and navigational queries. Two types of features have been used for the query-goal identification, namely user-click behavior and anchor-link distribution leading to an accuracy of roughly 90\%. 

A follow-up work from Liu et al.~\cite{liu2006} extends~\cite{lee2005} by extracting two additional features from the click-through data: \textit{n Clicks Satisfied} (nCS) evidence and \textit{top n Results Satisfied} (nRS). Based on the extended set of features and the decision tree model, this approach achieved 76\% precision and 66.7\% recall. 

Baeza-Yates et al.~\cite{yates2006} applied both supervised and unsupervised learning to detect the user intents among 3 categories: informational, not informational or ambiguous. They found that supervised learning shows better performance in most cases, however unsupervised learning can compliment the supervised learning in some cases. Through a combined approach, they were able to reach 55\% precision and 45\% recall. 

Jansen et al.~\cite{jansen2008} built a decision tree that utilizes the features extracted based on the query terms and the Web documents viewed by the user to classify queries. For the classes of queries, they adopted Broder's~\cite{broder2002} taxonomy and extended it to three hierarchy levels. This approach results in 75\% accuracy. 

Hu et al.~\cite{hu2009} proposed a probability approach that maps query intent to Wikipedia concepts. The probability of a query belongs to a concept is computed based on the distribution of query terms in Wikipedia articles of each category. This approach reached 93\% precision and 91\% recall for the queries in travel, personal name and job categories. 

Kathuria et al.~\cite{kathuria2010} classified web queries into informational, navigational or transactional using k-means clustering. Considered features include query length, the number of page views of the search engine results page and the number of query modifications. Results indicate that this approach is able to improve the accuracy of the classification by 15\% compared to the approach proposed by Jansen et al.~\cite{jansen2008}.

Kravi et al.~\cite{kravi2016one} argued that search tasks involving more complex information needs are more likely to be associated with multi-click queries. The authors proposed to use supervised classification to identify multi-clicked queries based on features extracted from the queries, clicks and search engine result page. They achieved 75.2\% accuracy for identifying multi-click queries.


The aforementioned works focus on the classification of single query sessions, often limited to data collected through lab studies. 
However, recent studies have shown that users information seeking tasks have grown more sophisticated~\cite{jones2008} and often require one or more queries across multiple search sessions~\cite{kotov2011,liu2010,agichtein2012}.
In contrast to these previous works, we propose to automatically detect the intent of search activities at the level of search missions (see Section \ref{sec:terms}). In this way, higher level information needs which are reflected in real-world search missions across multiple queries are considered, involving a larger amount of user activities as well as behavioral features are used for building supervised models.
\section{Problem}\label{sec:problem}
In this section, we first introduce the important terms used in this paper (Section \ref{sec:terms}), and then formally define the problem of detecting the intent of Web search missions from user interaction features (Section \ref{sec:definition}).

\subsection{Important Terms}\label{sec:terms}

Search sessions have been studied and defined by previous works from different perspectives~\cite{jones2008, jansen2007, jansen2003, silverstein1999}. To better understand the dialog between users and search engines, Hagen et al.~\cite{hagen2013} proposed to distinguish between:
\begin{itemize}
 \item\textbf{Physical Search Sessions}. Physical sessions are determined by the time gap between queries. A physical session starts when a user entered the first query and ends when either he/she leaves the search engine or there was a certain period of inactivity. A physical session, however, is not equal to a task, as within a single physical session a user may perform several tasks.
 \item\textbf{Logical Search Sessions}. Logical sessions are characterized by consecutive queries towards the same information need within the same physical session. Whenever there is a topic shift between two successive queries, the current logical session ends and a new logical session begins. Depending on the number of tasks a user is trying to perform, a physical session may contain one or more logical sessions.
 \item\textbf{Search Missions}. Search missions are characterized by groups of logical sessions that share the same information need. The logical sessions need not to be contiguous but may be interleaved within one or multiple physical sessions. 
 
\end{itemize}


\begin{table}[ht]
\caption{Example query log divided into physical search sessions, logical search sessions and search missions.}
\label{tab:sessionexample}
\centering
\begin{adjustbox}{max width=0.49\textwidth}
\begin{tabular}{ c l c c c c}
\toprule
{} & & & & \multirow{2}{3.5em}{\textbf{Logical Session}} & \multirow{2}{3.5em}{\textbf{Physical Session}}\\
{\textbf{ID}} & \multicolumn{1}{c}{\textbf{Query}} & \textbf{Timestamp} & \textbf{Mission} & &  \\ 
\midrule
1 & ancient turkey & 2012-12-20 20:02:44  & \multirow{2}{1.4em}{M1}& \multirow{2}{1.1em}{L1} & \multirow{2}{1.2em}{P1} \\ 
2 & history istanbul & 2012-12-20 20:24:17 \\ \\ \cdashline{1-6} \\
3 & istanbul archeology & 2012-12-21 12:02:54 & M1 & L2 & P2 \\  \\\cdashline{1-6} \\
4 & istanbul archeology & 2012-12-21 18:31:21  & M1 & L3& \multirow{8}{1.4em}{P3}\\ \\ \cdashline{1-5} \\
5 & weather new york & 2012-12-21 18:45:23 & M2&L4  & \\\\\cdashline{1-5}\\
6 & constantinople & 2012-12-21 18:45:36 &M1 & L5 & \\\\ \cdashline{1-5}\\
7 & footbal lisbon & 2012-12-21 19:14:01& \multirow{2}{1.4em}{M3} & \multirow{2}{1.4em}{L6} & \\ 
8 & football lisbon & 2012-12-21 19:14:11 \\ \\\cdashline{1-6}\\
9 & benfica vs sporting & 2012-12-21 20:23:04 & M3 & L7 & P4\\\\ \cdashline{1-6}\\
10 & derby eterno & 2012-12-21 22:42:48 & \multirow{3}{1.4em}{M1}& \multirow{3}{1.4em}{L8} & \multirow{3}{1.4em}{P5}  \\ 
11 & constantinople & 2012-12-21 23:09:02 \\
12 & constantinople & 2012-12-21 23:27:38 \\ 
\bottomrule
\end{tabular}
\end{adjustbox}
\end{table}

Table ~\ref{tab:sessionexample} shows an example query log divided into physical search sessions, logical search sessions and search missions. For instance, one of the search missions \textit{M1} consists of 5 logical sessions (\textit{L1, L2, L3, L5, L8}) that spread across 4 physical sessions (\textit{P1, P2, P3, P5}). 


Sessions and missions can be classified according to Broder's ``Taxonomy of Web Search''~\cite{broder2002}, which has been widely used in the Web search context. The author suggested that the traditional notion of an information need might not be adequate in describing Web search, and the ``need behind the query'' is often not informational in nature. Therefore, Broder classifies Web search queries according to the goal of the search task into three classes:

\begin{itemize}
 \item\textbf{Navigational}. The immediate intent is to reach a particular site. For instance, the query \textit{``Greyhound Bus''} targets the website \textit{http://www.greyhound.com}.
 
 \item\textbf{Informational}. The intent is to acquire some information assumed to be present on one or more Web pages. A typical example of an informational query is \textit{``What is a search engine?''}, where the purpose of this query is to find information assumed to be available on the Web and no further interaction is intended except reading.

 \item\textbf{Transactional}. The intent is to perform some Web-mediated activity. The main categories for such queries are shopping, finding various web-mediated services, downloading files (images, songs, etc) or the like. 
\end{itemize}

In later works~\cite{rose2004, yates2006}, Broder's taxonomy has been further studied and each category has been split into several subcategories. In this work, we adopt the original 3-class Broder's taxonomy for the search mission intent detection.

\subsection{Problem Definition} \label{sec:definition}


In this work, we consider a query log $M$ consisting of individual search missions. The log of an individual search mission $m_j \in M$ consists of: 
\begin{itemize}
\item \textit{queries:} a set of $n > 0$ queries $Q = \{q_1, q_2, \dots, q_n\}$ and the corresponding timestamps $T_Q = \{ t_{q1}, t_{q2}, \dots, t_{qn}\}$ indicating the query execution time for queries in $Q$. 

 
\item \textit{clicks:} a set of $k \geq 0$ URLs $U(q_i) = \{u_1, u_2, \dots, u_k\}$ of the Web documents clicked from the search engine result page (SERP) of each query $q_i \in Q$. Along with their clicking time $T_U(q_i) = \{t_{u1}, t_{u2}, \dots, t_{uk}\}$ and their ranks $R_{U}(q_i) = \{r_{u1}, r_{u2}, \dots, r_{uk}\}$ in the SERP.
\end{itemize}


On this basis, we define the task of this work as follows:
\begin{definition}{Detecting the Intent of Web Search Missions:}
Given a particular search mission $m_j$ log, we aim at classifying the search intent of $m_j$ into one of the three categories: informational, navigational and transactional.
\end{definition}

In particular, we aim at classifying missions by utilizing a range of user interaction features obtainable from any arbitrary query log consisting of the aforementioned elements

The focus of this work is to automatically detect the intent of Web search missions based on user behaviour and interaction within the mission. While we assume pre-labeled search missions in order to focus on the classification problem, research on the mission segmentation problem~\cite{hagen2013} is out of the scope.

Further details on how class labels are generated is given in Section \ref{sec:approach}.
\section{Approach} \label{sec:approach}
We approach the problem of detecting informational Web search missions with supervised models for classification. In this section, we describe the features (Section \ref{sec:feature}) as well as the classifiers (Section \ref{sec:classifier}) we used in the approach.

\subsection{Features}\label{sec:feature}
\begin{table*}[!htbp]
\caption{Features used for detecting the search intent of Web search missions.}
\label{tab:feature}
\centering
\begin{adjustbox}{width={0.9\textwidth}}
\begin{tabular}{cll}
\toprule
   \textbf{Type} & \textbf{Feature} & \textbf{Description}\\
\midrule
   \multirow{7}{*}{Query} & $q\_min$ & minimum number of query terms among all queries in the mission\\
   & $q\_max$ & maximum number of query terms among all queries in the mission\\
   & $q\_avg$ & average number of query terms of all queries in the mission\\
   & $q\_unique$ & total number of unique query terms of all query in the mission\\
   & $q\_cos3$ & average cosine similarity of 3-grams between two consecutive queries in the mission\\
   & $q\_cos4$ & average cosine similarity of 4-grams between two consecutive queries in the mission\\
   & $q\_lehv$ & average Lehvenstein distance between two consecutive queries\\
   \midrule
   \multirow{6}{*}{Mission} & $m\_queries$ & number of queries in the mission\\
   & $m\_logical$ & number of logical sessions in the mission\\
   & $m\_duration+break$ & mission duration including break duration between consecutive logical sessions\\
   & $m\_duration-break$ & mission duration excluding break duration between consecutive logical sessions\\
   & $m\_avg+break$ & average mission duration including break duration between consecutive logical sessions per query ($\frac{m_duration+break}{m\_queries}$)\\
   & $m\_avg-break$ &  average mission duration excluding break duration between consecutive logical sessions per query  ($\frac{m_duration-break}{m\_queries}$)\\
   \midrule
   \multirow{7}{*}{Browsing} & $b\_click$ & total number of clicked Web documents from all SERPs in the mission\\
   & $b\_unique$ & total number of unique domains visited throughout mission\\
   & $b\_revisit$ & total number of revisited domains\\
   & $b\_revisitunique$ & number of unique revisited domains\\
   & $b\_clickrate$ & ratio of queries in the mission correspond to at least one click on its SERP\\
   & $b\_cos3$ & average cosine similarity of 3-grams between each query and one of the corresponding clicked domain URLs pairs\\
   & $b\_cos4$ & average cosine similarity of 4-grams between each query and one of the corresponding clicked domain URLs pairs\\
   & $b\_avg\_SERPs$ & average number of SERP visits per query\\
   & $b\_SERPs$ & total number of SERP visits in the mission\\
\bottomrule
\end{tabular}
\end{adjustbox}
\end{table*}

We extract features according to multiple dimensions of a search mission, structured into three categories, namely features related to a \textit{Query}, \textit{Mission} and \textit{Browsing} behaviour. 

All considered features are listed in Table \ref{tab:feature}. The details of the features and the intuition behind them is described in detail in the remainder of this section.

\subsubsection{Query-based features.} 
Previous studies by Lee et al.~\cite{lee2005} and Liu et al.~\cite{liu2006} found that navigational queries are typically short in length whereas informational queries are longer~\cite{kathuria2010}. Studies on search as learning also found that after being exposed with information, users tend to reconstruct their queries that are more in line to what they are searching for~\cite{eickhoff2014lessons}. This is especially true when the searchers are not familiar with the domain itself. On the other hand reformulation in navigational search sessions tends to be minimal, and is often due to a spelling mistake. With a simple word/character edit distance, this behavior could potentially be detected. These findings motivate the extraction of features related to \textit{number of query terms} ($q\_min$, $q\_max$, $q\_avg$, $q\_unique$) and the \textit{between query similarity} ($q\_cos3$, $q\_cos4$, $q\_lehv$) in this category.


\subsubsection{Session-based features.} 
Studies by Kathuria et al.~\cite{kathuria2010} have shown that the informational search sessions are longer because users usually spend longer time on viewing documents and search engine result pages (SERPs). Sessions with transactional intent often span over multiple physical and logical sessions. For instance, when planning for a holiday, each single mission involved here, such as researching destinations or the booking of travel tickets and hotels, can last for several logical sessions.
Based on these findings, we assume that the session-based features, i.e. \textit{total number of queries issued ($m\_queries$), \textit{mission duration} ($m\_duration+break$, $m\_duration-break$), \textit{mission duration per query} ($m\_avg+break$, $m\_avg-break$) and the \textit{number of logical sessions} ($m\_logical$)}, are intuitively effective for distinguishing between navigational missions and informational/transactional missions.

\subsubsection{Browsing-based features.} 
Lee et al.~\cite{lee2005} and Liu et al.~\cite{liu2006} exploited search result click-through data to automatically identify users' search intent. The assumption is that with navigational queries, most clicks are alloted to just one results position, thus the \textit{number of clicks} ($b\_click$, $b\_unique$) is much less than informational queries. Kravi et al.~\cite{kravi2016one} also found that search with complex information needs is more often correlated with multi-click queries. Thus we assume that the features related to \textit{number of clicks on SERPs ($b\_click$, $b\_unique$), number of visits to SERPs ($b\_SERPs$, $b\_avg\_SERPs$)} are effective for identifying the type of intent. More detailed features
corresponding to the browsing behaviour have also been studied by Gwizdka et al.~\cite{gwizdka2006can}, indicating that the more complex a task is for a user, the higher the ratio of \textit{revisited pages ($b\_revisit$, $b\_revisitunique$)} . 
Furthermore, due to the nature of navigational and transactional missions, the \textit{similarity between query and the clicked URL ($b\_cos3$, $b\_cos4$)} in the missions tends to be higher than it is in the informational missions. 


\subsection{Supervised Classification of Missions} \label{sec:classifier}
We aim at learning a supervised classification model that is able to classify a search mission into one of the three classes based on the introduced feature sets. For the classification model, we have experimented with several different approaches. Considering the scale of the data (Section \ref{sec:experiment}) as well as the number and characteristics of the features, we have opted for Decision Tree (DT), Logistic Regression (LR), Support Vector Machine (SVM)~\cite{platt1999} and Random Forest (RF)~\cite{breiman2001} as classification models. We tune the hyper parameters of each classifier through grid search. The performance of the best configurations of each classifier is reported in Section \ref{sec:evaluation}.

\section{Experimental Setup}\label{sec:experiment}

\subsection{Data}
For experimentally evaluating our approach, we rely on the dataset\footnote{http://www.uni-weimar.de/medien/webis/corpora/corpus-webis-smc-12/corpus-webis-smc-12.zip} created by Hagen et al.~\cite{hagen2013}, which consists of 8840 web search queries from 127 users. The search log has been segmented in sessions and missions by two experts, resulting in 2881 logical sessions and 1378 missions. On average, each user fired 69.6 queries spread across 22.7 logical sessions. Each mission consists of 6.5 queries across 2 logical sessions.

\subsection{Ground Truth}

In order to create the ground truth for the experimental evaluation of our search intent detection approach, we manually labeled all the missions in the aforementioned dataset. 
To ensure the quality of the labeling result, we assign each search mission to two different annotators. Each annotator inspected the entire log of a mission and assigned a label describing its search intent referring to one of the three classes in Broder's Taxonomy: informational, navigational and transactional. If the intent can not be inferred explicitly, the mission was labeled as ``ambiguous''. 

The level of agreement between the annotators (see Table~\ref{tab:agreement}) shows that the manual classification process can be difficult even for human annotators. Together the annotators had a 66\% inter-rater agreement (913 missions: 454 informational, 275 navigational and 184 transactional respectively). The remaining ones are either labeled as ``ambiguous'' or labeled differently by the two annotators and have been removed from the ground truth dataset in order to maintain a high quality ground truth. Our final ground truth dataset contains 6860 queries from 124 users corresponding to 2136 logical sessions and 913 missions. 

\begin{table}[!htb]
\caption{Agreement table of the ground truth labeling result.}
\label{tab:agreement}
\centering
\begin{adjustbox}{max width=0.49\textwidth}
\begin{tabular}{l|cccc}
\toprule
\diagbox{\textbf{Label 1}}{\textbf{Label 2}} & \textbf{Informational} & \textbf{Navigational} & \textbf{Transactional} & \textbf{Ambiguous}\\ \midrule
\textbf{Informational} & \textbf{454} & 14 & 74 & 29\\
\textbf{Navigational} & 64 & \textbf{275} & 46 & 30\\
\textbf{Transactional} & 58 & 10 & \textbf{184} & 18\\
\textbf{Ambiguous} & 46 & 2 & 9 & 65\\
\bottomrule
\end{tabular}
\end{adjustbox}
\end{table}

\subsection{Baselines \& Configurations}

\subsubsection{Configurations}\label{sec:config}

We report the performance of the different configurations of our mission classification approach as listed below:

\begin{itemize}
\item \textit{Classifiers.} We apply a range of standard models for the classification, namely, Decision Tree (DT), Logistic Regression (LR), Support Vector Machine (SVM)~\cite{platt1999} and Random Forest (RF)~\cite{breiman2001}. For our experiments, we used the open source machine learning workbench WEKA\footnote{https://www.cs.waikato.ac.nz/ml/index.html}~\cite{witten2016data,hall2009weka}. 
We tune the hyper parameters for best overall accuracy with grid search to optimize the result, the detailed results of different hyper parameters is omitted in this paper. 
In Section \ref{sec:evaluation}, we report the result of the best performing hyper parameter configuration for each classifier.

\item \textit{Balanced vs. unbalanced training data.} As shown in Table \ref{tab:agreement}, the number of instances in different classes are not equally distributed in our ground truth dataset, which might affect the performance of the classifiers. In order to find the best performing configuration of our approach, we experimented with both \emph{unbalanced} (i.e. real world distribution) and \emph{balanced} training sets.


\end{itemize}

\subsubsection{Baseline}
Our approach, reflected in the configurations described above, is based on the assumption that the search intent of a user is best studied when considering entire search missions, potentially involving one or more logical or physical sessions. For this reason, we consider the classification of logical sessions as baseline and compare the classification performance in longer units with more user activities (missions) with smaller units with less user activities (logical sessions).

\subsubsection{Evaluation Metrics}

We run 10-fold cross validation on all the configurations as described in Section \ref{sec:config}, and evaluate the results according to the following metrics:

\begin{itemize}
\item \textbf{Accuracy ($Accu$):} percentage of search missions that were classified with the correct class label according to our ground truth. 
\item \textbf{Precision ($P$), Recall ($R$), F1 ($F1$) score of class $i$:} the standard precision, recall and F1 score on the classification result of each class $i$.
\item \textbf{Weighted average of precision ($P$), recall ($R$), and F1 ($F1$):} the weighted average of the corresponding score across all 3 classes. The weights proportional to class frequencies in the dataset.
\end{itemize}

Further, to analyze the usefulness of individual features, we make use of the \textit{Information Gain (IG)} metric, which is measured as the reduction of entropy (uncertainty) regarding the classification of the test class based on the observation of a particular feature. 
\section{Evaluation Results}\label{sec:evaluation}
This section reports the experimental evaluation result based on the setup as described in Section \ref{sec:experiment} and a preliminary analysis of the feature importance.


\subsection{Classification Performance}


\begin{table*}[!htbp]
	\caption{Performance of different classifiers (mission classification).}
	\label{tab:eva_classifier}
	\small
	\centering
 	\scalebox{1.3}{
	\begin{tabular}{p{0.5cm}p{1cm}| p{0.5cm}p{0.5cm}p{0.5cm}|p{0.5cm}p{0.5cm}p{0.5cm}|p{0.5cm}p{0.5cm}p{0.5cm}|p{0.5cm}p{0.5cm}p{0.5cm}|p{0.5cm}p{0.5cm}p{0.5cm}|p{0.5cm}}
	\toprule
   & & \multicolumn{3}{c}{\textbf{Navigational}}& \multicolumn{3}{c}{\textbf{Informational}}& \multicolumn{3}{c}{\textbf{Transactional}}& \multicolumn{3}{c}{\textbf{Weighted average}} & \textbf{All} \\
  & \textbf{Method} & \textbf{P}& \textbf{R}& \textbf{F1}& \textbf{P}& \textbf{R}& \textbf{F1}& \textbf{P}& \textbf{R}& \textbf{F1}& \textbf{P}& \textbf{R}& \textbf{F1} &\textbf{Accu} \\
   \toprule
   
\parbox[t]{2mm}{\multirow{4}{*}{\rotatebox[origin=c]{90}{\textbf{unbalanced}}}} & DT & 0.764 & 0.731 & 0.747 & 0.644 & 0.839 & 0.728 & 0.241 & \textbf{0.076} & \textbf{0.116} & 0.599 & 0.653 & 0.611 & 0.653\\
 & SVM & 0.786 & \textbf{0.760} & \textbf{0.773} & \textbf{0.656} & 0.927 & 0.768 & \textbf{0.800} & 0.022 & 0.042 & \textbf{0.724} & \textbf{0.694} & 0.623 & \textbf{0.694}\\
 & LR & \textbf{0.809} & 0.709 & 0.756 & 0.651 & \textbf{0.938} & \textbf{0.769} & 0.556 & 0.054 & 0.099 & 0.680 & 0.691 & \textbf{0.630} & 0.691\\
 & RF & 0.782 & 0.731 & 0.756 & 0.648 & 0.923 & 0.761 & 0.556 & 0.027 & 0.052 & 0.670 & 0.685 & 0.617 & 0.685\\

   \bottomrule
   \toprule
   \parbox[t]{2mm}{\multirow{4}{*}{\rotatebox[origin=c]{90}{\textbf{balanced}}}} & DT & 0.740 & 0.753 & 0.746 & 0.473 & 0.652 & 0.548 & 0.441 & 0.266 & 0.332 & 0.551 & 0.557 & 0.542 & 0.557\\
 & SVM & 0.757 & 0.785 & 0.771 & 0.470 & \textbf{0.725} & 0.570 & 0.477 & 0.201 & 0.283 & 0.568 & 0.570 & 0.541 & 0.570\\
 & LR & \textbf{0.779} & \textbf{0.789} & \textbf{0.784} & \textbf{0.499} & 0.676 & \textbf{0.574} & \textbf{0.507} & \textbf{0.321} & \textbf{0.393} & \textbf{0.595} & \textbf{0.595} & \textbf{0.584} & \textbf{0.600}\\
 & RF & 0.745 & 0.756 & 0.751 & 0.484 & 0.652 & 0.555 & 0.435 & 0.277 & 0.339 & 0.555 & 0.562 & 0.548 & 0.562 \\

   \bottomrule
	\end{tabular}}
\end{table*}

Here we report the evaluation result of the different configurations of our approach. The results are shown in Table \ref{tab:eva_classifier}. For all configurations, the classification accuracy is above 0.557 and the F1 score is above 0.541, which indicates that the set of features we
extracted from user search activities can provide meaningful evidence for detecting the search intent. A more detailed discussion is presented in the remaining sections.
\subsubsection{Performance on unbalanced training data.}
The performance of different classifiers using unbalanced training data is shown in the upper part of Table \ref{tab:eva_classifier}. 
The results indicate that SVM (0.694) and LR (0.691) have better performance with respect to overall accuracy compared to DT (0.653) and RF (0.685). A similar trend can also be observed with respect to the weighted average F1 score across classes, where SVM (0.623) and LR (0.630) outperform DT (0.611) and RF (0.617). As for precision, the highest precision of all classes are achieved by either SVM or LR, which suggest that any approach aiming at high precision should best resort to SVM or LR.

However, when considering the classification performance of each individual class, all classifiers have better performance in terms of precision, recall and F1 score on the navigational and informational class than on the transactional class. Despite the fact that SVM results in a reasonable precision, all classifiers failed to recall more than 92\% of the transactional missions. This might be caused by the unbalanced distribution of classes in the ground truth dataset, as well as the ambiguous nature of the transactional missions. We will further discuss this observation in Section \ref{sec:eva_balanced}. 




\subsubsection{Performance on balanced training data.} \label{sec:eva_balanced}
The performance of different classifiers using balanced training data is shown in the lower section of Table \ref{tab:eva_classifier}. 

In terms of the overall accuracy, LR (0.600) and SVM (0.570) again outperform RF (0.562) and DT (0.557). The highest F1 score (0.584) is also achieved by the LR classifier. In terms of precision, the highest score of each individual class as well as the weighted average across classes are all achieved by the LR classifier. This implies that, LR is the best option from the investigated classifiers for high precision applications when the training set is balanced.

When considering the performance of classifiers on each individual class, the highest precision, recall and F1 score are all achieved for the navigational class. All metrics on the informational class show poorer performance compared to the unbalanced set. On the other hand, the metrics on the transactional class increases after balancing the training set. This suggest that the informational missions and the transactional missions are very often showing similar characteristics on the features we extracted, hence are easy to be confused with each other.

Despite the fact that the performance increases on the transactional class, it still appears as the most challenging class among the three. The poor performance of all classifiers on the transactional class indicate that this class is very ambiguous, i.e. transactional search intent is more difficult to detect compared to the other two classes. We infer that more focused and class-specific approaches are required to improve performance for particular classes.

When comparing between the overall performance of classifiers on balanced and unbalanced data, the results indicate that the best overall accuracy is 0.094 percentage points higher on the unbalanced dataset. Despite the navigational class showing similar results on both configurations, the performance on the informational class is better when using unbalanced training data whereas the performance on the transactional class is better when using balanced training data. This implies that the training setup should be designed with a specific application in mind. For instance, for the task of detecting learning-related search missions, that is, informational missions, the classifiers should be trained on the dataset following its natural distribution (i.e. unbalanced data), to achieve high performance.  

\subsection{Baseline Comparison}

\begin{figure*}[!htbp]
\centering
	\subfloat[b][Average F1 score using unbalanced training data]{\includegraphics[clip=true, trim=2pt 0pt 4pt 4pt, width=0.47\textwidth]{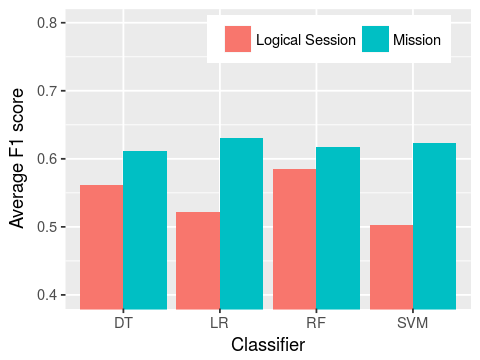}\label{fig:f1_unb}} 
	\subfloat[b][Accuracy using unbalanced training data]{\includegraphics[width=0.47\textwidth]{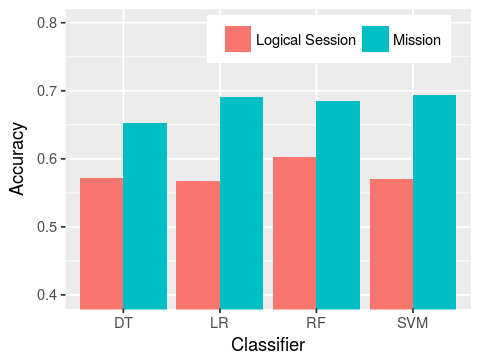}\label{fig:accuracy_unb}}	\newline
    	\subfloat[b][Average F1 score using balanced training data]{\includegraphics[width=0.47\textwidth]{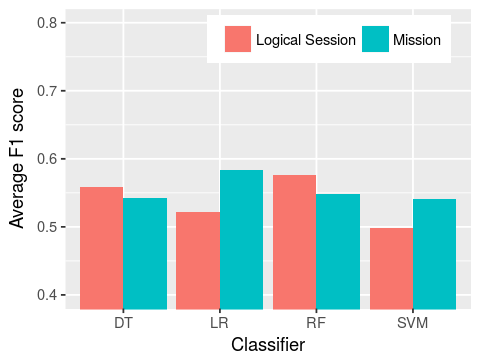}\label{fig:f1_b}} 
	\subfloat[b][Accuracy using balanced training data]{\includegraphics[width=0.47\textwidth]{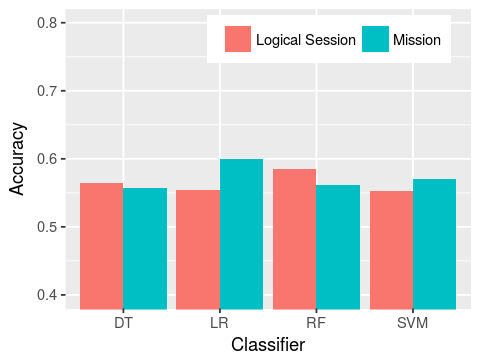}\label{fig:accuracy_b}}
    
    \caption{Average F1 score and accuracy of the search intent classification at different units (mission vs. logical session).}\label{fig:units}
\end{figure*}

We compared the performance of our approach, classifying entire search missions, and baselines utilising the signals from logical sessions only. The result of using both unbalanced (Figure \ref{fig:f1_unb}, \ref{fig:accuracy_unb}) and balanced (Figure \ref{fig:f1_b}, \ref{fig:accuracy_b}) training data is shown in Figure \ref{fig:units}. From the figure we observe that the performance using missions is consistently better than for logical sessions across all models when using unbalanced training data, considering both the average F1 score and the overall accuracy. The highest average F1 score and overall accuracy are also achieved on missions when using balanced training data. The best performance in both cases is achieved by the LR classifier, where the average F1 score for mission classification outperforms the logical session by 0.046 (0), and the accuracy by 0.091 (0.014) using unbalanced (balanced) training set. This result supports our assumption that, by investigating queries within the context of a whole mission, one can gather more signals and consequently, identify the user's intent more precisely. 

From the comparison result we observe that the average F1 and accuracy gap between mission and logical session drops 0.062 and 0.095 respectively on average after balancing the training data. The main cause is the significant performance decrease of the mission classification after balancing the training data. One of the potential reasons behind this is that the balancing of the training data reduces the weight of the larger class (i.e. informational class) in the training process. As the detection of the informational class is challenging in nature, high performance is achieved through a large number of instances. With the reduction of training instances implied by the balancing step, the performance of the classifier decreases on this class. While the logical session classification task has more training instances than the missions classification task in our case, the performance decrease on logical sessions is less significant than on missions. 

\subsection{Feature Impact}

\begin{figure*}[!htbp]
\centering
\includegraphics[clip=true, trim=2pt 0pt 4pt 10pt, width=0.96\textwidth]{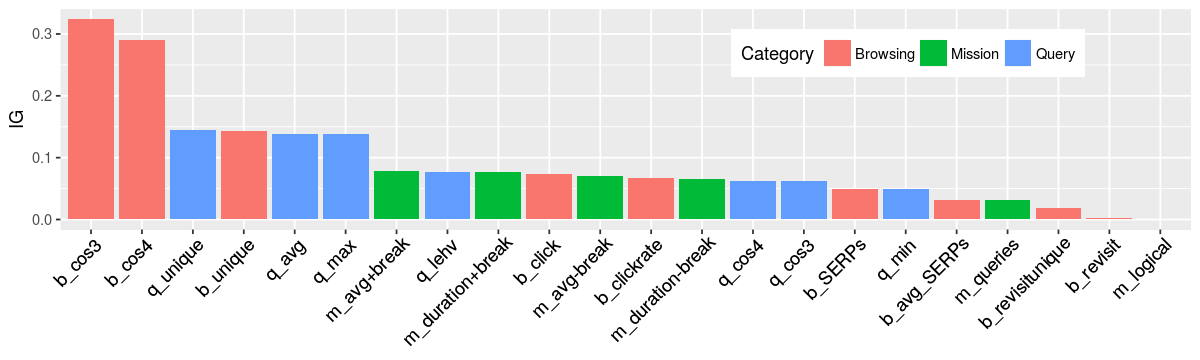}
\caption{Information gain of features used for classification.}
\label{fig:feature_ig}
\end{figure*}

The information gain of each feature is plotted in Figure \ref{fig:feature_ig}. 
We observe that the features which show the highest information gain are the features utilising the cosine similarity between the query and the clicked domain URL ($b\_cos3$ and $b\_cos4$). This is intuitive since these features are strong indicators for differentiating between navigational queries, where a very high similarity is expected, and the other two intents. 
While there is a significant gap between these two features and subsequent ones according to their information gain ranking, subsequently ranked features show less significant differences with respect to their contribution to the classification result. 
The next-ranking feature is the number of unique terms of all queries in the mission ($q\_unique$) which reflects the complexity of the information need to a certain extent. That is, in navigational search, the number of unique terms usually remains constant, whereas in informational and transactional search the number of unique terms is likely to increase as the search session grew longer. 
The feature ranking forth is the number of unique domains clicked by the user ($b\_unique$), which is powerful in differentiating navigational queries from the rest due to the fact that navigational queries have specific targets which supposedly are easier by users to identify. 
The features reflecting average and max query length ($q\_avg$, $q\_max$) have been studied by previous works and shown to be correlated with the complexity of information need, and are ranked fifth and sixth. The remaining features beyond the first six ranks do have only minor contributions on the classification result with the bottom feature having information gain of 0.

Overall, based on the IG result, features in the browsing category appear more important than features in other categories, with 2 browsing features ranking at the top 2 positions. Query features are also shown to be effective with 3 features among the top 6. Mission-based features have the least contribution among all 3 categories. This seems intuitive, given that all three types of missions usually (a) involve multiple queries and clicks, (b) can have long browsing time once the user found the Web documents satisfying the information need and (c) could consist of several logical sessions, which makes the mission-related features less effective.
\section{Discussion}
In this work, we explore the possibility of detecting search intent based on user behavior and interactions observed during the actual search process. Whereas the overall classification performance indicates reasonable results on average (Section \ref{sec:evaluation}), with SVM and LR providing the best performance on average, in particular transactional missions appear ambiguous for both human annotators as well as supervised models. For this reason, results indicate that more specific classification tasks are likely to yield superior performance. For instance, an application-specific classifier aimed at targeted advertising may focus on only transactional or informational missions (depending on the advertised offering), so that binary classification can be applied through a more tailored model.

Similarly, further optimisation might be geared towards either recall or precision, depending on the application goal. As indicated by the results, even on supposedly hard task of detecting transactional missions, SVM and LR appear to provide reasonable precision, yet comparably low recall, indicating that high precision-classification might be achievable at the cost of recall even for challenging classification goals. 

Limitations of our work in particular arise from the characteristics of the experimental dataset~\cite{hagen2013}. While search engines have evolved significantly throughout the last years, so has the user search behavior. In particular, given the highly personalised as well as supervised approaches shown by state-of-the-art search engines, users have to apply less effort as well as time during search missions. Hence, supervised models require recent training data in order to be applicable to current search behavior.

In addition, the choice of features used in this work was limited by the data available in the experimental dataset. However, further features have been shown to be effective for the classification task at hand. For instance, Arguello et al.~\cite{arguello2014predicting} discovered that informational tasks are associated with longer time spent on SERPs and higher mouse activity. However, due to the lack of availability of such data, our model was constrained to the features presented in the previous sections.

\section{Conclusions}
In this paper, we propose a supervised approach for classifying the search intent of a user during Web search missions. In particular, we are referring to three established categories, namely navigational, informational and transactional missions. Based on the study of previous works and the observation of search session data, we propose and extract 22 features from an experimental dataset based on a real-world online query log. Based on a expert-labeled ground truth, our performance results suggest that features capturing the user interaction and behavior provides a sound basis for supervised classification of search missions. Potential applications of this work are manifold, for instance, as part of improved retrieval and ranking of search results, to better facilitate learning and support information needs during Web search or to recommend additional information during online search missions, for instance, as part of targeted online advertising or resource recommendations. 

Since limitations arise in particular from the nature of the experimental dataset and the lack of publicly available, up-to-date query logs, future work will be concerned in particular with the application of similar approaches on a more recent and larger scale dataset. This would enable supervised models which are better reflecting contemporary search behavior and at the same time, utilise a wider variety of features.

\bibliographystyle{ACM-Reference-Format}
\bibliography{reference} 
\balance
\end{document}